\documentclass[preprint2]{aastex}

\newcommand{\myemail}{sdzib@mpifr-bonn.mpg.de}
\newcommand{\msec}[2]{$#1\mbox{$''\mskip-7.6mu.\,$}#2$}
\newcommand{\rahms}[4]{$#1^{\rm h}#2^{\rm m}#3\mbox{$^{\rm s}\mskip-7.6mu.\,$}#4$} 
\newcommand{\decdms}[4]{$#1^{\circ}#2'#3\mbox{$''\mskip-7.6mu.\,$}#4$} 

\slugcomment{First Draft}

\shorttitle{Kinematic of the Trapezium--BN/KL region}
\shortauthors{Dzib et al.}

\usepackage{epsfig}
\usepackage{longtable}
\usepackage{amssymb}
\usepackage{lscape}
\usepackage{booktabs}

\begin{document}

\newpage

\title{RADIO MEASUREMENTS OF THE STELLAR PROPER MOTIONS IN THE CORE OF THE ORION NEBULA CLUSTER}

\author{Sergio A. Dzib\altaffilmark{1}, 
Laurent Loinard\altaffilmark{1,2}, 
Luis F.\ Rodr\'{\i}guez\altaffilmark{2,3},
Laura G\'omez\altaffilmark{4,5,6},
Jan Forbrich\altaffilmark{7,8},
Karl M. Menten\altaffilmark{1},
Marina A. Kounkel\altaffilmark{9},
Amy J.\ Mioduszewski\altaffilmark{10},
Lee Hartmann\altaffilmark{9},  
John J. Tobin\altaffilmark{11,12} and
Juana L.\ Rivera\altaffilmark{2}
}

\altaffiltext{1}{Max Planck Institut f\"ur Radioastronomie, Auf dem H\"ugel 69, 53121 Bonn, Germany (\myemail)}

\altaffiltext{2}{Instituto de Radioastronom\'{\i}a y Astrof\'{\i}sica, Universidad
Nacional Aut\'onoma de M\'exico\\ Apartado Postal 3-72, 58090,
Morelia, Michoac\'an, Mexico }

\altaffiltext{3}{El Colegio Nacional, Donceles 104, 06020, M\'exico, DF, M\'exico}
\altaffiltext{4}{Joint ALMA Observatory, Alonso de C\'ordova 3107, Vitacura, Santiago, Chile}
\altaffiltext{5}{CSIRO Astronomy and Space Science, PO Box 76, NSW 1710 Epping, Australia}
\altaffiltext{6}{Departamento de Astronom\'\i a, Universidad de Chile, Camino El Observatorio 1515, Las Condes, Santiago, Chile}
\altaffiltext{7}{University of Vienna, Department of Astrophysics, Türkenschanzstr. 17, 1180 Vienna, Austria}
\altaffiltext{8}{Harvard-Smithsonian Center for Astrophysics, 60 Garden St, Cambridge, MA 02138, USA}
\altaffiltext{9}{Department of Astronomy, University of Michigan, 500 Church Street, Ann Arbor, MI 48105, USA}
\altaffiltext{10}{National Radio Astronomy Observatory, Domenici Science Operations Center,
1003 Lopezville Road, Socorro, NM 87801, USA}
\altaffiltext{11}{Homer L. Dodge Department of Physics and Astronomy, University of Oklahoma, 440 W.
Brooks Street, Norman, OK 73019, USA}
\altaffiltext{12}{Leiden Observatory, Leiden University, P.O. Box 9513, 2300 RA Leiden, The Netherlands }

\begin{abstract}
Using multi--epoch VLA observations, covering a time baseline  of 29.1 years, we
have measured the proper motions of 88 young stars with compact radio emission in the 
{core of the Orion Nebula Cluster (ONC)} and the neighboring BN/KL region. Our work increases the number of
young stars with measured proper motion at radio frequencies by a factor of 2.5 and 
enables us to perform a better statistical analysis of the kinematics of the region
than was previously possible. Most stars (79 out of 88) have proper motions consistent
with a Gaussian distribution centered on $\overline{\mu_{\alpha}\cos{\delta}}=1.07\pm0.09\quad{\rm mas\,yr^{-1}}$,
and $\overline{\mu_{\delta}}=-0.84\pm0.16\quad{\rm mas\,yr^{-1}}$, with velocity dispersions
of $\sigma_{\alpha}=1.08\pm0.07\quad{\rm mas\,\,yr^{-1}},$ $\sigma_{\delta}=1.27\pm0.15\quad{\rm mas\,\,yr^{-1}}$. 
We looked for organized movements of these stars but found no clear indication of
radial expansion/contraction or rotation. The remaining nine stars in our sample
show peculiar proper motions that differ from the mean proper 
motions of the ONC by more than 3\,$\sigma$. 
One of these stars, V~1326 Ori, could have been expelled from the
Orion Trapezium 7,000 years ago. Two could be related to the multi-stellar disintegration 
in the BN/KL region, in addition to the previously known sources BN, I and n. 
The others either have high uncertainties (so their anomalous proper
motions are not firmly established) or could be foreground objects.

\end{abstract}


\section{INTRODUCTION}

The angular resolution of radio interferometers improves proportionally 
with the longest baseline in the array.  With baselines of tens of kilometers, 
as with the Karl G. Jansky Very Large Array (VLA), angular resolutions of order \msec{0}{1} 
are possible around $\nu$  = 10 GHz. This enables source positions 
to be measured to about \msec{0}{01} even for moderate signal-to-noise detections.
In addition, interferometric radio observations
are usually phase--referenced with respect to background quasars whose positions are
accurately measured in the International Celestial Reference Frame (ICRF).
Thus, although they are not strictly absolute, the positions delivered by 
radio interferometers at multiple epochs can be directly  compared, and accurate 
proper motions can be measured. 
Indeed, VLA observations with time separations of several years have
been used to measure proper motions with errors of the order of 
1.0~mas~yr$^{-1}$ \citep{2003ApJ...583..330R, 2002RMxAA..38...61L, 
2014ApJ...788..162D}. 

A significant fraction of Young Stellar Objects (YSOs) are
radio emitters thanks to two main mechanisms: 
thermal bremsstrahlung (free-free) and non-thermal gyrosynchrotron emission. 
Both classes can be observed with the VLA. Thus, multi-epoch VLA observations 
can be used to accurately measure the proper motions of YSOs. This is particularly 
interesting because, unlike optical or near-infrared observations, radio measurements 
are essentially immune to obscuration by dust.

The Orion Nebula Cluster (ONC) is the nearest region 
\citep[d=414$\pm7$ pc,][]{2007A&A...474..515M}\footnote{Recent results of the Gould's Belt 
Distances survey ({\sc GOBELINS}) suggest a smaller distance of 388$\pm$5 pc 
(Kounkel et al. 2016a, submitted). { This new value does not affect our results significantly.}}
having recently formed
massive stars. In the {core} of the ONC (Hillenbrand \& Hartmann 1998), there are two sub--regions 
of particular interest: the Orion Trapezium and the Orion BN/KL region, which is located behind the ONC. 
Together they cover an area of a few square arcminutes.
This area is also  one of the most crowded with young 
stellar objects emitting at radio frequencies 
\citep[i.e.,][]{1987ApJ...314..535G, 1987ApJ...321..516C, 2004AJ....127.2252Z, 2014ApJ...790...49K, 2016ApJ...822...93F}. 
Taking advantage of these characteristics and using multi-epoch observations, 
the proper motion of several objects were measured by \cite{2005ApJ...635.1166G, 2008ApJ...685..333G}.
Here we will expand and improve this study by adding recent observations
made with the newly expanded VLA
at similar frequencies and resolutions as those used by \cite{2005ApJ...635.1166G, 2008ApJ...685..333G}.
We will use the new proper motions to study the overall Galactic motion 
of the cluster, its internal kinematics, and to identify radio sources with peculiar
velocities. Finally, our study can be compared with kinematic studies at optical
wavelengths of radial velocities (Kounkel et al. 2016b) and proper motions
(Poveda et al. 2005).

\section{OBSERVATIONS}

We searched the VLA archive for deep ($<$150 $\mu$Jy  beam$^{-1}$) 
observations recorded in the most
extended A configuration at C band (4 to 8 GHz) 
and X band (8 to 12 GHz). The frequencies and configuration
were chosen as a compromise between angular resolution (better
than 0$\rlap{.}\,''$5) and the area covered by the field of view.
These two bands are also the most sensitive of the VLA.

Observations with these characteristics were found for
the following epochs (see also Table \ref{tab:Ep}): 1985 January 19, 1995 July 2, %
2000 November 13, 2006 May 12 (the observations up to this date were previously
analyzed by \citealt{2005ApJ...635.1166G, 2008ApJ...685..333G}), 2011 July 2, 2011 July 24,
2011 August 29 (results of these observation were
reported by \citealt{2014ApJ...790...49K}), 2012 October 3 \citep{2016ApJ...822...93F},\,
and 2014 March 3 (archival data reported here for the first time).
All these observations used the quasar J0541-0541 as  phase calibrator,
a source that is at an angular separation of $1\rlap{.}^\circ6$ from the ONC core.
We include an additional epoch, 1991 September 6, 
to more uniformly cover the time baseline of $\sim$30 years
between the first and last observed date. However, 
the quasar J0501-019 was used as phase calibrator in this
epoch. The angular separation of this quasar from the ONC core is $9\rlap{.}^\circ1$.
Such a large separation will affect the measured positions for this epoch 
(e.g., Pradel et al. 2006).
In fact, G\'omez et al.~(2008) estimated a systematic offset
in declination of $\Delta\delta=-0\rlap{.}\,''035$ for this observation,
which we will use here as well.

Observations that were taken prior to the VLA expansion in 2010
were calibrated, edited and imaged in AIPS as described in G\'omez et al. (2005, 2008).
The remaining epochs were processed similarly using
the CASA software (McMullin et al. 2007). Positions of the sources were obtained using 
a two dimensional gaussian fit (task 'imfit' in CASA).
For epochs 2011.36, 2011.50 and 2011.56, in order to 
obtain a better noise level than that reported by Kounkel et al. (2014),
we combined the two recorded sub-bands (each 1 GHz wide and centered, 
respectively, at 4.5 and 7.5 GHz). Also, when a given source was detected in several of the 
three epochs reported by Kounkel et al.\ (2014), we use a weighted average 
of the positions, and a time-stamp corresponding to the weighted mean of
the corresponding epochs.

\begin{table}[ht!]
\footnotesize
  \begin{center}
  \caption{Trapezium-BN/KL observations and final parameter of maps.}\label{tab:Ep}
    \begin{tabular}{cccc}\hline\hline
  & $\lambda$    &  Synthesized beam  & rms noise \\
Epoch  &(cm) &$\theta_{\rm maj}['']\times\theta_{\rm min}[''];$ P.A.[$^\circ$] & ($\mu$Jy bm$^{-1}$)\\
   \noalign{\smallskip}
\hline\hline\noalign{\smallskip}
\noalign{\smallskip}
\hline\noalign{\smallskip}
1985.05 & 6.0 &$0.43\times 0.35;\, -15$    & 136  \\
1991.67 & 3.6 &$0.26\times 0.25;\, -55$    & 77  \\
1995.56 & 3.6  &$0.26\times 0.22;\, +34$    & 42  \\
2000.87 & 3.6 &$0.24\times 0.22;\, +3$  & 40 \\
2006.36 & 3.6 &$0.26\times 0.22;\, -2$   & 58 \\ 
2011.50 & 5.0 &$0.30\times 0.27;\, +46$ & 115 \\
2011.56 & 5.0 &$0.47\times 0.24;\, -47$  & 110 \\
2011.66 & 5.0 &$0.33\times 0.25;\, -30$  & 102 \\  
2012.76 & 4.0 &$0.22\times 0.20;\, -7$  & 14 \\
2014.17 & 5.5 &$0.44\times 0.29;\, -38$  & 100 \\
2014.17 & 3.3 &$0.25\times 0.18;\, -38$  & 45\\
 \hline\hline\noalign{\smallskip}
\end{tabular}
\end{center}
\end{table}

\begin{figure*}[ht!]
\begin{center}
\includegraphics[width=01.0\textwidth,trim= 30 0 40 30, clip]{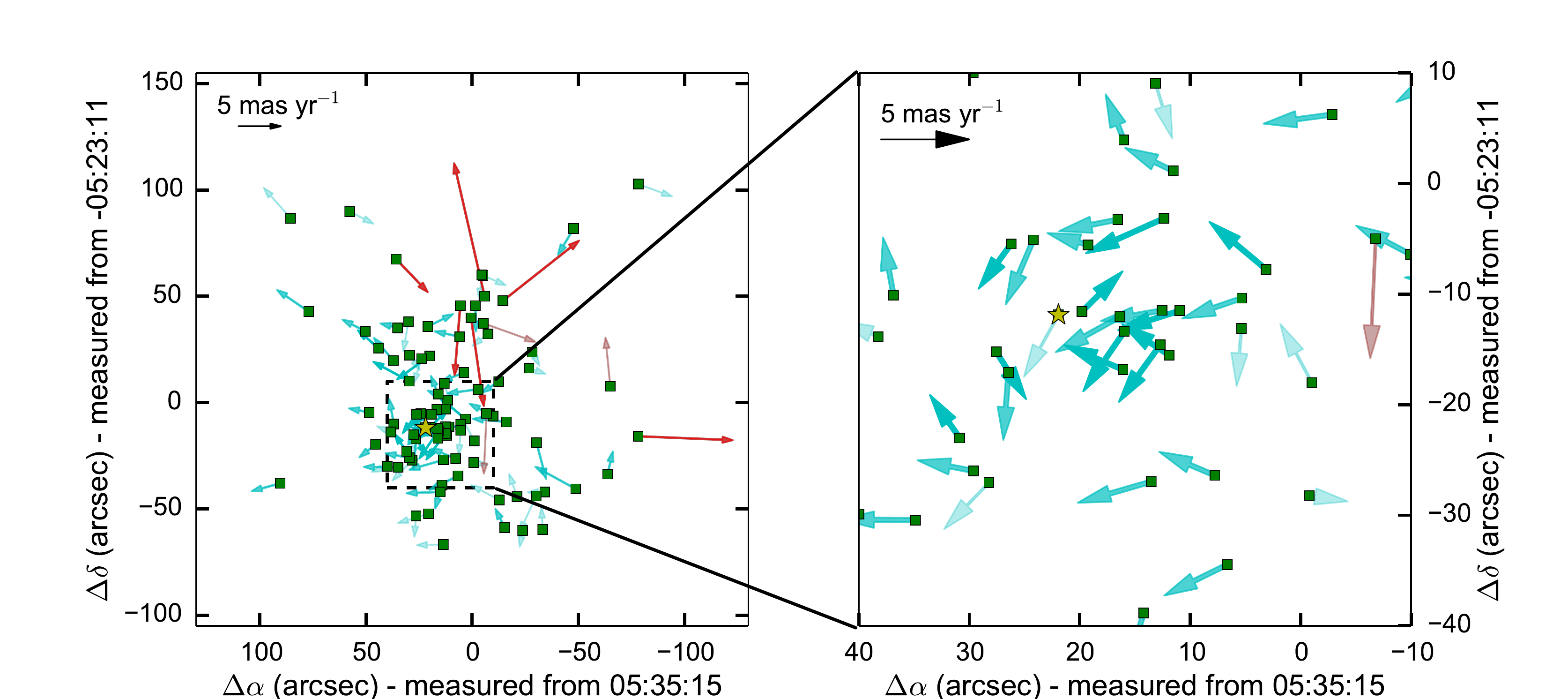}
\end{center}
\caption{Absolute proper motions of radio stellar sources  as 
distributed on the plane of the sky. Cyan arrows
are proper motions that are within 3 sigma of the mean proper
motions (see text) and red arrows correspond to those
sources above this limit. The yellow star indicates the position of the
massive star $\theta^{1}$ Ori C.  The level of transparency of the arrows indicates
the significance of the measured proper motions.
Left: the complete sample of measured proper motions.
Right: a zoom to the central region. }
\label{fig:pm}
\end{figure*}

\section{RESULTS}

After measuring the positions of the sources in 
individual epochs, we used them to compute
their proper motions. In order to obtain accurate results we restricted our proper
motion measurements to those radio sources with at 
least three detections and a minimum separation between
the first and the last detected epochs of 5 years. 
A total of 92 sources fulfill these requirements. This is more than
twice than those previously analyzed by G\'omez et al. (2005).
From these sources, four are related to the explosive
event in the BN/KL region and will be discussed in a
separate paper by Rodr\'{\i}guez et al. (2016).
Proper motions for the remaining 88 sources were obtained using
a least--squares fitting to the positions of the sources. 
Systematic errors of the order of 10~mas were added to 
the errors given by imfit, in order to obtain a reduced
$\chi^{2}$ of one.  The systematic errors are expected to be dominated by uncertainties
introduced by the interpolation of the phase calibration from the quasar to the targets field.
The measured proper motions
are listed in columns 3 and 4 of Table~\ref{tab:PM} and their distribution in
the plane of the sky are shown in Figure~\ref{fig:pm}. From our sample, 
for 14 stars (16\%) the proper motion measurements are significant above 2-$\sigma$
in both directions (right ascension and declination), for 43 (49\%) the proper motion in at least
one direction is above 2-$\sigma$, while for 31 (35\%) the measurements
are below 2-$\sigma$ in both directions. To highlight the first two cases in Figure~\ref{fig:pm}
we used different levels of transparency when we draw the arrows.
To calculate the transverse velocities corresponding to these proper motions,
we use 1 mas yr$^{-1}\equiv\,$1.96 km s$^{-1}$ as appropriate for the distance (414 pc;
Menten et al. 2007) to the ONC.

\begin{figure*}[ht!]
\begin{center}
\begin{tabular}{cc}
\includegraphics[width=0.45\textwidth]{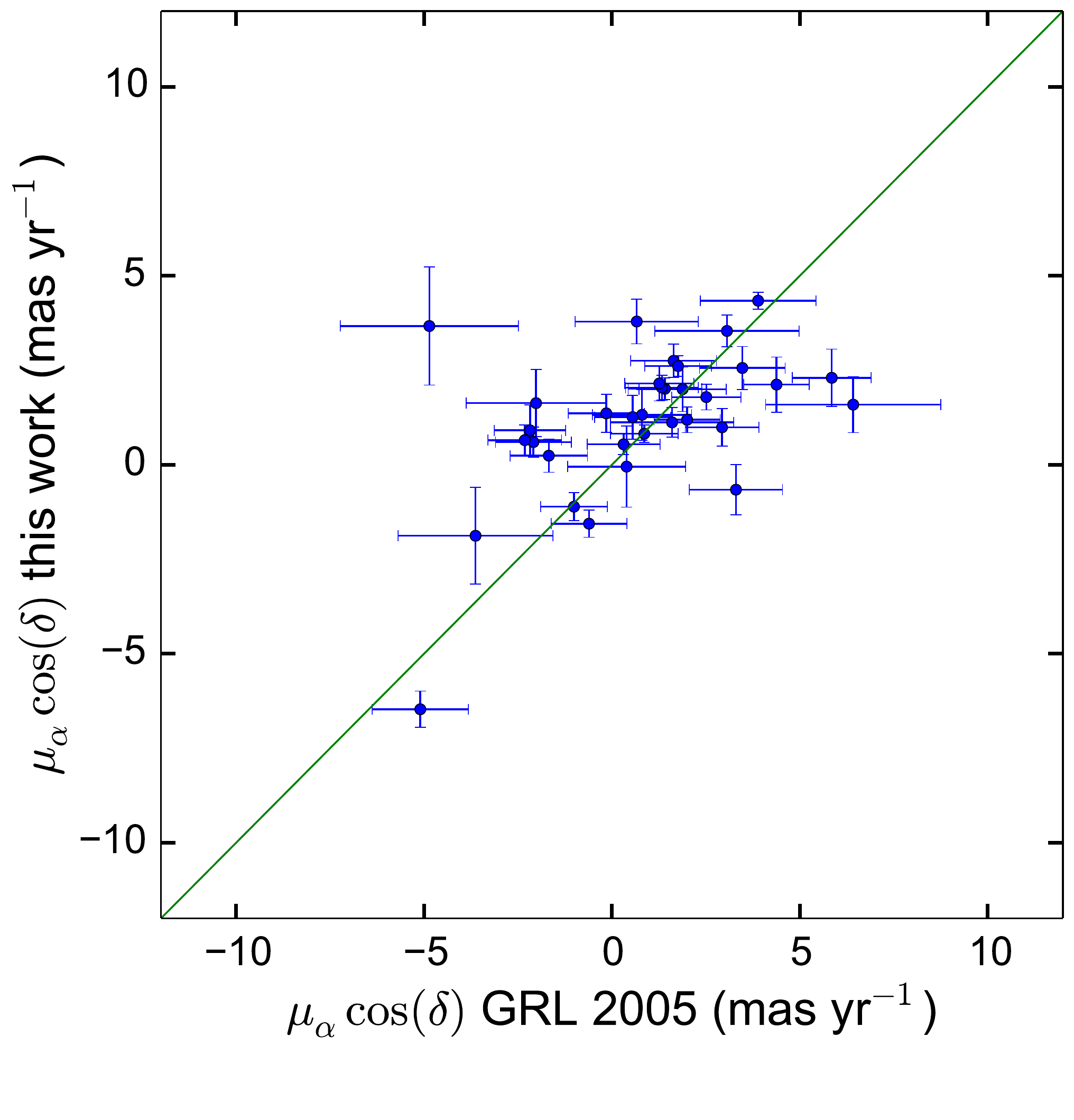}&
\includegraphics[width=0.45\textwidth]{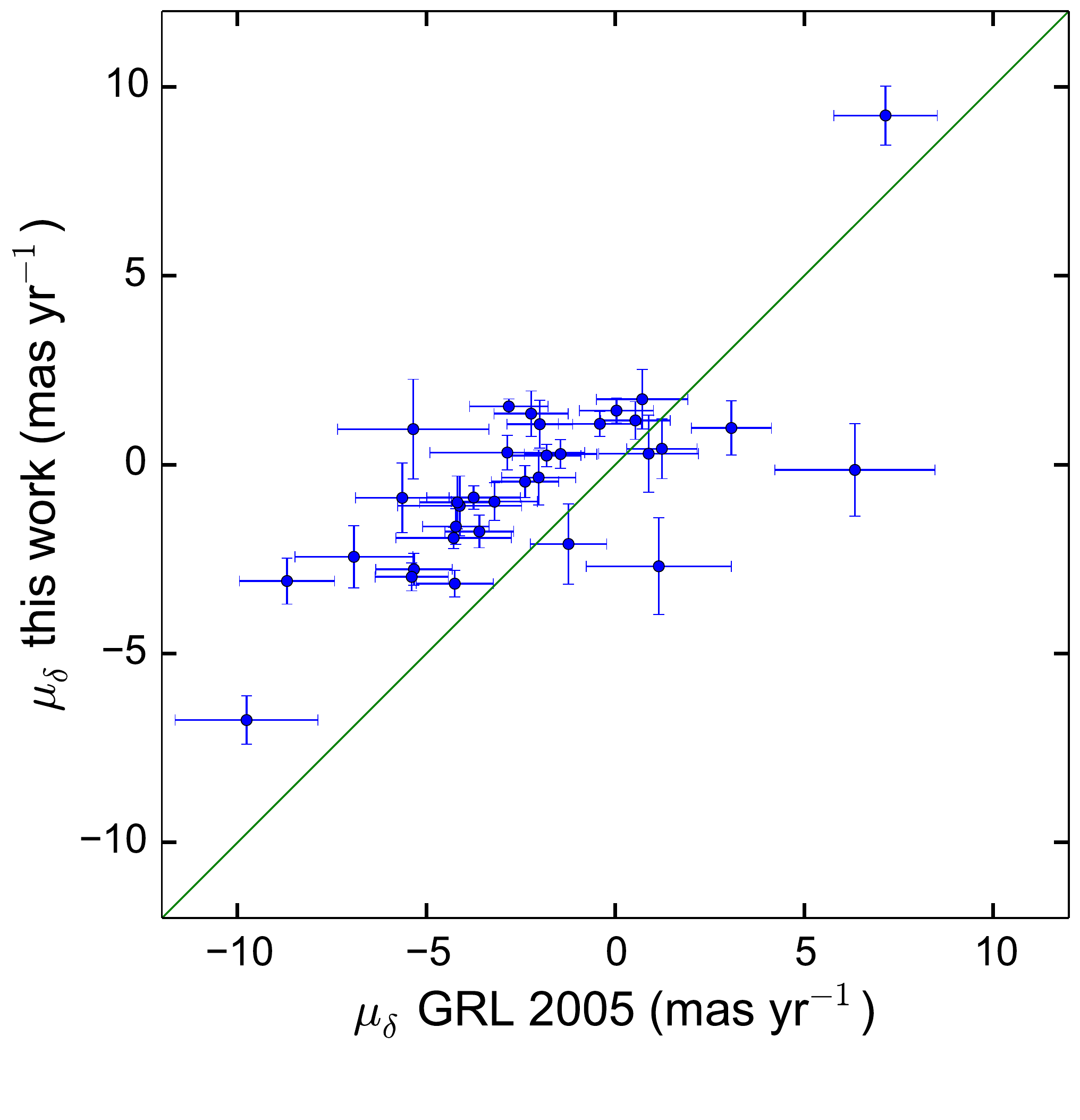}\\
\includegraphics[width=0.45\textwidth]{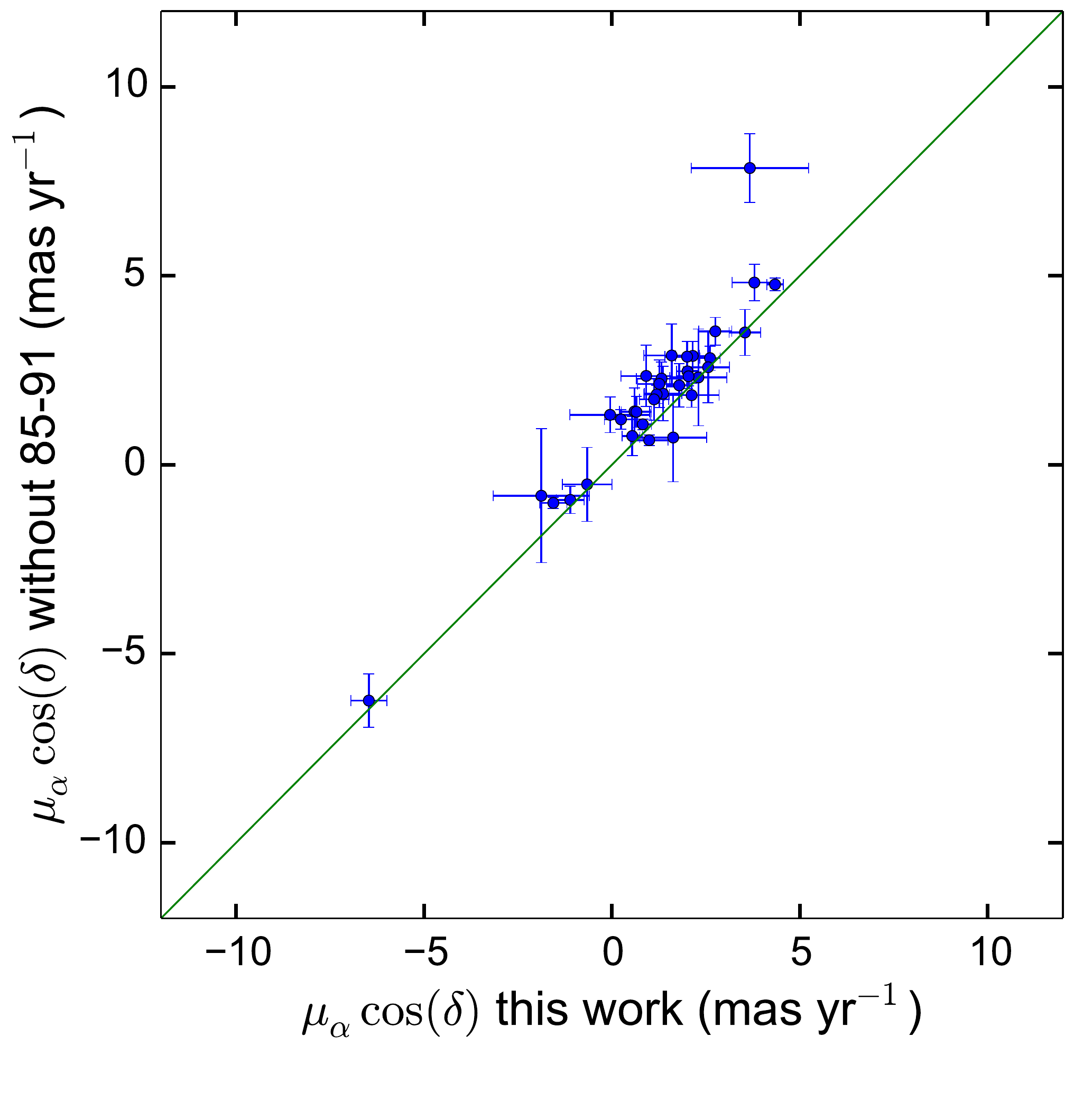}&
\includegraphics[width=0.45\textwidth]{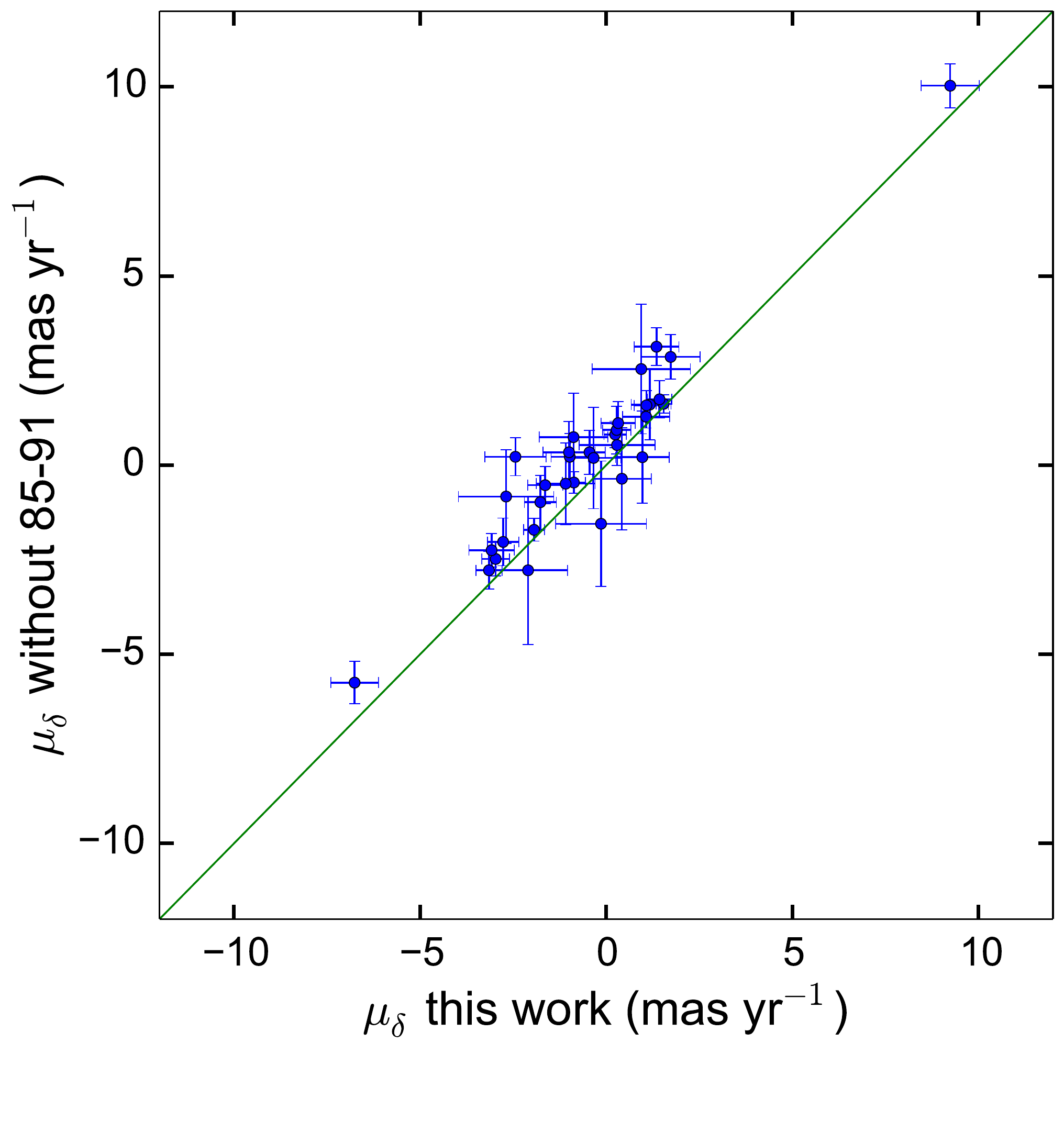}\\
\end{tabular}
\end{center}
\caption{The top panels show our measured proper motions against the proper motions measured 
by G\'omez et al. (2005; GRL 2005). The bottom panels show our measured proper motions with
all epochs considered against our measured proper motions without epochs 1985.05 and 1991.68.
The left panels are for right ascension and the right panels for 
declination.}
\label{fig:comp}
\end{figure*}

We now compare our results with previous results. In the top panels
of Figure \ref{fig:comp}, we show the 
comparison between our measured absolute proper motion and those obtained
by G\'omez et al. (2005) for the 35 radio sources that we have in common. 
While most of the proper motions in 
right ascension agree within 1$\sigma$, in declination
there is a systematic shift. This systematic shift can be attributed
to the different quasar used in the 1991.68 epoch. As we mentioned before, 
G\'omez et al. (2008) found a systematic shift in declination of 
$\Delta\delta=-0\rlap{.}\,''035$  that was not used
by G\'omez et al. (2005). To test our hypothesis we measured the proper
motion of the same 35 sources without epochs 1985.05 and 1991.68 and compare 
them with the proper motions from Table~\ref{tab:PM}, this comparison
is shown in the bottom panels of Figure \ref{fig:comp}. In this case we see 
that the vast majority of the proper motions agree in both right ascension 
and declination. Thus, we confirm that it is correct to use the $\Delta\delta$ 
value found by G\'omez et al. (2008) for the epoch 1991.68.

\begingroup
\let\clearpage\relax
\begin{deluxetable}{cccccc}
\tabletypesize{\scriptsize}
\tablewidth{0pt}
\tablecolumns{8}
\tablecaption{PROPER MOTIONS OF THE RADIO SOURCES IN ORION\label{tab:PM}}
\tablehead{           &  &\multicolumn{2}{c}{Absolute}&\multicolumn{2}{c}{Orion's rest frame}  \\
\cmidrule(lr){3-4}\cmidrule(lr){5-6}          & Other &$\mu_\alpha \cos{\delta}$ & $\mu_\delta$ & $\mu_\alpha \cos{\delta}$ & $\mu_\delta$  \\
\colhead{VLA Name} & \colhead{Name\tablenotemark{a}} &\colhead{(mas yr$^{-1}$)} & \colhead{(mas yr$^{-1}$)}& \colhead{(mas yr$^{-1}$)} & \colhead{(mas yr$^{-1}$)}\\
}\startdata
J053509.76-052128.2&COUP 342&-3.77$\pm$2.96&-1.39$\pm$3.85&-4.84$\pm$2.96&-0.55$\pm$3.85\\
J053509.77-052326.9&V1326 Ori&-13.23$\pm$5.17&-0.54$\pm$1.96&-14.3$\pm$5.17&0.3$\pm$1.97\\
J053510.65-052303.4&COUP 391&0.59$\pm$1.48&6.0$\pm$3.08&-0.48$\pm$1.48&6.84$\pm$3.08\\
J053510.73-052344.6&COUP 394&-0.46$\pm$0.93&1.96$\pm$0.97&-1.53$\pm$0.93&2.8$\pm$0.98\\
J053511.74-052351.6&COUP 443&4.92$\pm$0.97&2.69$\pm$1.63&3.85$\pm$0.97&3.53$\pm$1.64\\
J053511.80-052149.2&GMR A&1.63$\pm$0.89&-2.69$\pm$1.28&0.56$\pm$0.89&-1.85$\pm$1.29\\
J053512.71-052353.1&\nodata&1.06$\pm$2.47&-0.02$\pm$4.3&-0.01$\pm$2.47&0.82$\pm$4.3\\
J053512.78-052410.7&COUP 496&0.11$\pm$1.64&1.83$\pm$2.4&-0.96$\pm$1.64&2.67$\pm$2.41\\
J053512.97-052330.0&COUP 509&-1.09$\pm$2.59&-4.1$\pm$1.81&-2.16$\pm$2.59&-3.26$\pm$1.82\\
J053512.98-052355.0&COUP 510&1.93$\pm$1.46&-4.2$\pm$5.64&0.86$\pm$1.46&-3.36$\pm$5.64\\
J053513.11-052247.3&[B2000] j131-247 &-0.37$\pm$2.02&-0.7$\pm$1.59&-1.44$\pm$2.02&0.14$\pm$1.6\\
J053513.21-052254.8&COUP 530&-1.08$\pm$0.65&-0.38$\pm$1.54&-2.15$\pm$0.66&0.46$\pm$1.55\\
J053513.41-052411.2&Zapata 10&0.17$\pm$0.88&-0.95$\pm$2.11&-0.9$\pm$0.88&-0.11$\pm$2.12\\
J053513.59-052355.3&COUP 554&0.69$\pm$1.03&2.01$\pm$1.16&-0.38$\pm$1.03&2.85$\pm$1.17\\
J053513.93-052320.1&COUP 593&3.64$\pm$1.47&0.34$\pm$0.78&2.57$\pm$1.47&1.18$\pm$0.8\\
J053513.97-052409.8&COUP 594&0.75$\pm$0.59&1.49$\pm$0.66&-0.32$\pm$0.6&2.33$\pm$0.68\\
J053514.03-052223.2&Zapata 11&-10.68$\pm$5.59&8.43$\pm$2.33&-11.75$\pm$5.59&9.27$\pm$2.34\\
J053514.14-052356.8&COUP 607&3.0$\pm$2.02&1.56$\pm$3.12&1.93$\pm$2.02&2.4$\pm$3.12\\
J053514.16-052301.1&GMR C&1.36$\pm$0.5&-0.88$\pm$0.92&0.29$\pm$0.51&-0.04$\pm$0.93\\
J053514.33-052317.4&COUP 625&2.28$\pm$0.87&1.2$\pm$1.23&1.21$\pm$0.87&2.04$\pm$1.24\\
J053514.50-052238.7&COUP 639&1.18$\pm$1.06&-0.91$\pm$0.99&0.11$\pm$1.06&-0.07$\pm$1.0\\
J053514.55-052316.0&COUP 640&0.36$\pm$2.09&-7.86$\pm$6.72&-0.71$\pm$2.09&-7.02$\pm$6.72\\
J053514.61-052221.0&IRc 23&4.43$\pm$3.72&19.29$\pm$8.64&3.36$\pm$3.72&20.13$\pm$8.64\\
J053514.65-052233.7&Parenago 1839&-6.63$\pm$7.1&-2.37$\pm$2.39&-7.7$\pm$7.1&-1.53$\pm$2.4\\
J053514.66-052211.2&COUP 647&-2.09$\pm$2.97&-0.9$\pm$1.37&-3.16$\pm$2.97&-0.06$\pm$1.38\\
J053514.69-052211.0&COUP 647&0.26$\pm$1.34&-2.14$\pm$1.42&-0.81$\pm$1.34&-1.3$\pm$1.43\\
J053514.81-052304.8&MLLA 472&3.24$\pm$0.73&-0.43$\pm$1.28&2.17$\pm$0.74&0.41$\pm$1.29\\
J053514.90-052225.4&GMR D&-0.05$\pm$1.07&-2.44$\pm$0.82&-1.12$\pm$1.07&-1.6$\pm$0.84\\
J053514.93-052329.0&COUP 671&1.44$\pm$1.68&2.67$\pm$1.72&0.37$\pm$1.68&3.51$\pm$1.73\\
J053514.95-052339.2&Parenago 1844&-0.54$\pm$1.96&-0.08$\pm$2.16&-1.61$\pm$1.96&0.76$\pm$2.17\\
J053515.03-052231.1&MLLA 606&-1.77$\pm$8.47&-12.23$\pm$3.75&-2.84$\pm$8.47&-11.39$\pm$3.75\\
J053515.21-052318.8&COUP 690&2.84$\pm$1.1&2.38$\pm$1.18&1.77$\pm$1.1&3.22$\pm$1.19\\
J053515.26-052256.9&Zapata 29&1.97$\pm$0.91&-0.81$\pm$0.98&0.9$\pm$0.91&0.03$\pm$0.99\\
J053515.36-052321.4&MLLA 410&2.58$\pm$0.93&-0.88$\pm$0.92&1.51$\pm$0.93&-0.04$\pm$0.93\\
J053515.36-052324.1&MLLA 391&0.19$\pm$1.0&-2.25$\pm$1.37&-0.88$\pm$1.0&-1.41$\pm$1.38\\
J053515.38-052225.4&MLLA 630C&0.75$\pm$1.24&-9.28$\pm$3.88&-0.32$\pm$1.24&-8.44$\pm$3.88\\
J053515.40-052240.0&[H97b] 20055&2.31$\pm$1.01&0.49$\pm$1.2&1.24$\pm$1.01&1.33$\pm$1.21\\
J053515.44-052345.5&Parenago 1868&3.04$\pm$0.84&-1.49$\pm$0.88&1.97$\pm$0.84&-0.65$\pm$0.89\\
J053515.52-052337.4&GMR 14&2.29$\pm$0.76&0.97$\pm$0.72&1.22$\pm$0.77&1.81$\pm$0.74\\
J053515.73-052322.5&GMR 26&2.75$\pm$0.44&-0.87$\pm$0.31&1.68$\pm$0.45&-0.03$\pm$0.35\\
J053515.77-052309.9&$\theta^{1}$ Ori E&1.67$\pm$0.39&0.8$\pm$0.58&0.6$\pm$0.4&1.64$\pm$0.6\\
J053515.80-052326.5&GMR 13&1.78$\pm$0.34&1.17$\pm$0.5&0.71$\pm$0.35&2.01$\pm$0.52\\
J053515.83-052314.1&GMR 12&4.34$\pm$0.22&-1.94$\pm$0.28&3.27$\pm$0.24&-1.1$\pm$0.32\\
J053515.84-052322.5&GMR 11&2.6$\pm$0.27&-0.45$\pm$0.42&1.53$\pm$0.28&0.39$\pm$0.45\\
J053515.85-052325.6&GMR 10&2.0$\pm$0.29&-2.76$\pm$0.42&0.93$\pm$0.3&-1.92$\pm$0.45\\
J053515.88-052301.9&COUP 743&-0.65$\pm$1.64&-2.14$\pm$1.61&-1.72$\pm$1.64&-1.3$\pm$1.62\\
J053515.91-052338.0&GMR 24&3.78$\pm$0.59&-1.09$\pm$0.78&2.71$\pm$0.6&-0.25$\pm$0.8\\
J053515.91-052417.8&COUP 748&2.52$\pm$1.89&-0.05$\pm$1.25&1.45$\pm$1.89&0.79$\pm$1.26\\
J053515.95-052349.8&GMR 9&0.6$\pm$0.4&-2.1$\pm$1.05&-0.47$\pm$0.41&-1.26$\pm$1.06\\
J053516.00-052353.0&Zapata 46&3.67$\pm$1.56&-0.15$\pm$1.22&2.6$\pm$1.56&0.69$\pm$1.23\\
J053516.07-052324.4&GMR 8&2.04$\pm$0.33&-2.97$\pm$0.37&0.97$\pm$0.34&-2.13$\pm$0.4\\
J053516.07-052307.0&GMR 15&0.54$\pm$0.27&1.35$\pm$0.6&-0.53$\pm$0.28&2.19$\pm$0.62\\
J053516.08-052327.8&GMR 22&2.56$\pm$0.57&1.07$\pm$0.33&1.49$\pm$0.58&1.91$\pm$0.37\\
J053516.10-052323.0&TCC 58&3.2$\pm$0.94&-1.76$\pm$1.69&2.13$\pm$0.94&-0.92$\pm$1.7\\
J053516.11-052314.3&TCC 59&2.39$\pm$0.93&-0.47$\pm$0.96&1.32$\pm$0.93&0.37$\pm$0.97\\
J053516.29-052316.6&GMR 7&0.82$\pm$0.23&0.24$\pm$0.3&-0.25$\pm$0.25&1.08$\pm$0.34\\
J053516.33-052322.6&GMR 16&-1.56$\pm$0.36&1.54$\pm$0.2&-2.63$\pm$0.37&2.38$\pm$0.26\\
J053516.34-052249.0&Zapata 54&3.78$\pm$0.78&-2.72$\pm$1.01&2.71$\pm$0.79&-1.88$\pm$1.02\\
J053516.38-052403.3&Parenago 1895&3.25$\pm$1.86&-0.99$\pm$1.49&2.18$\pm$1.86&-0.15$\pm$1.5\\
J053516.40-052235.2&GMR K&-2.51$\pm$0.89&1.24$\pm$1.5&-3.58$\pm$0.89&2.08$\pm$1.51\\
J053516.47-052322.9&$\theta^{1}$ Ori C&1.68$\pm$1.4&-3.01$\pm$1.94&0.61$\pm$1.4&-2.17$\pm$1.95\\
J053516.59-052250.3&MLLA 532&0.76$\pm$1.68&-2.34$\pm$1.39&-0.31$\pm$1.68&-1.5$\pm$1.4\\
J053516.62-052316.1&GMR 21&1.32$\pm$0.67&-3.08$\pm$0.61&0.25$\pm$0.68&-2.24$\pm$0.63\\
J053516.75-052316.5&GMR 6&1.19$\pm$0.34&-1.64$\pm$0.47&0.12$\pm$0.35&-0.8$\pm$0.5\\
J053516.77-052404.3&V1279 Ori&0.13$\pm$0.81&-1.66$\pm$2.47&-0.94$\pm$0.82&-0.82$\pm$2.48\\
J053516.77-052328.1&GMR 17&0.24$\pm$0.44&-3.14$\pm$0.35&-0.83$\pm$0.45&-2.3$\pm$0.38\\
J053516.85-052326.2&GMR 5&-1.11$\pm$0.37&-1.76$\pm$0.43&-2.18$\pm$0.38&-0.92$\pm$0.46\\
J053516.89-052338.1&Zapata 62&1.98$\pm$1.83&-2.05$\pm$1.71&0.91$\pm$1.83&-1.21$\pm$1.72\\
J053516.97-052248.7&GMR E&-0.66$\pm$0.66&-0.98$\pm$0.5&-1.73$\pm$0.67&-0.14$\pm$0.52\\
J053516.98-052337.0&GMR 4&2.15$\pm$0.46&0.41$\pm$0.79&1.08$\pm$0.47&1.25$\pm$0.81\\
J053516.98-052300.9&COUP 845&4.2$\pm$1.23&2.26$\pm$0.87&3.13$\pm$1.23&3.1$\pm$0.88\\
J053517.01-052233.0&V1333 Ori&0.53$\pm$2.84&-3.01$\pm$2.58&-0.54$\pm$2.84&-2.17$\pm$2.58\\
J053517.07-052334.0&GMR 3&0.99$\pm$0.49&1.43$\pm$0.33&-0.08$\pm$0.5&2.27$\pm$0.37\\
J053517.33-052341.4&[H97b] 20009&3.76$\pm$0.69&0.04$\pm$1.3&2.69$\pm$0.7&0.88$\pm$1.31\\
J053517.35-052235.9&GMR L&1.12$\pm$0.39&0.29$\pm$1.02&0.05$\pm$0.4&1.13$\pm$1.03\\
J053517.39-052203.6&MLLA 712&-3.81$\pm$1.88&-3.98$\pm$2.39&-4.88$\pm$1.88&-3.14$\pm$2.4\\
J053517.47-052321.1&COUP 885&0.42$\pm$1.06&2.29$\pm$1.02&-0.65$\pm$1.06&3.13$\pm$1.03\\
J053517.48-052251.2&[H97b] 20031&2.0$\pm$0.89&2.32$\pm$1.61&0.93$\pm$0.89&3.16$\pm$1.62\\
J053517.56-052324.9&GMR 2&0.65$\pm$0.4&0.27$\pm$0.38&-0.42$\pm$0.41&1.11$\pm$0.41\\
J053517.68-052340.9&GMR 1&0.91$\pm$0.67&-0.34$\pm$0.72&-0.16$\pm$0.68&0.5$\pm$0.74\\
J053517.95-052245.4&GMR G&2.0$\pm$0.58&1.73$\pm$0.79&0.93$\pm$0.59&2.57$\pm$0.81\\
J053518.05-052330.7&GMR 19&1.26$\pm$0.58&-1.0$\pm$0.7&0.19$\pm$0.59&-0.16$\pm$0.72\\
J053518.24-052315.6&Zapata 75&1.59$\pm$0.74&0.31$\pm$0.46&0.52$\pm$0.75&1.15$\pm$0.49\\
J053518.37-052237.4&GMR F&2.12$\pm$0.72&1.07$\pm$0.63&1.05$\pm$0.73&1.91$\pm$0.65\\
J053518.86-052141.2&Parenago 1924&-2.26$\pm$2.84&-1.16$\pm$1.84&-3.33$\pm$2.84&-0.32$\pm$1.85\\
J053520.15-052228.2&COUP 1084&3.6$\pm$1.6&2.46$\pm$1.45&2.53$\pm$1.6&3.3$\pm$1.46\\
J053520.72-052144.3&V1239 Ori&3.15$\pm$3.59&3.54$\pm$4.09&2.08$\pm$3.59&4.38$\pm$4.09\\
\enddata
\tablenotetext{a}{ COUP = Getman et al. (2005), V = Kukarkin et al. (1971), GMR = Garay et al. (1987), 
[B2000] = Bally et al. (2000), Zapata = Zapata et al. (2004), IRc = Rieke et al. (1973), Parenago = Parenago (1954),
MLLA = Muench et al. (2002), and [H97b] = Hillenbrand (1997).}
\vspace{-0.6cm}
\end{deluxetable}

\endgroup

From the same top panels of Figure \ref{fig:comp}, an improvement in the error
can be noticed between the old and new measurement of proper motions. These 
improvements are on average of the order of two.
As can be seen from Appendix A, the proper motion errors are expected to 
decrease with time as {t}$^{-3/2}$, in ideal cases. Thus, by duplicating the 
time baseline we expect an improvement in errors by a factor of 2.8.
Differences between the real improvements and those expected can be due to
the non-uniform sampling when observing and to different position errors between 
the observed epochs.

Menten et al. (2007) used the Very Long Baseline
Array (VLBA) telescope to observe four non--thermal YSOs (GMR~A, GMR~12, GMR~G
and GMR~F) and obtained accurate astrometry for them. Our VLA astrometry for 
these sources is consistent to better than 1.5 $\times\sigma$ with that reported
by Menten et al. (2007).
The only exception is for the right ascension component of the proper motion of GMR G,
for which the VLBA value is about twice the figure obtained with the VLA.  
Many VLBA--detected YSOs belong to tight multiple systems (Ortiz-Le\'on et al.~2016 submitted).
In some systems a close companion could cause differences of measured
proper motions between the VLA and the VLBA, since the former will measure the total 
motion of the system and the latter could be dominated by an orbital motion component
(e.g., Loinard et al. 2007).  
More recent VLBA observations of GMR~G by Kounkel et al. (2016a) 
obtained different values for the proper motions than those reported by Menten 
et al. (2007) suggesting that the motion of the source is not uniform, perhaps due 
to a close companion.

\section{ANALYSIS AND DISCUSSION}

\subsection{Mean proper motions and velocity dispersions}

\begin{figure*}[t!]
\begin{center}
\includegraphics[width=0.80\textwidth]{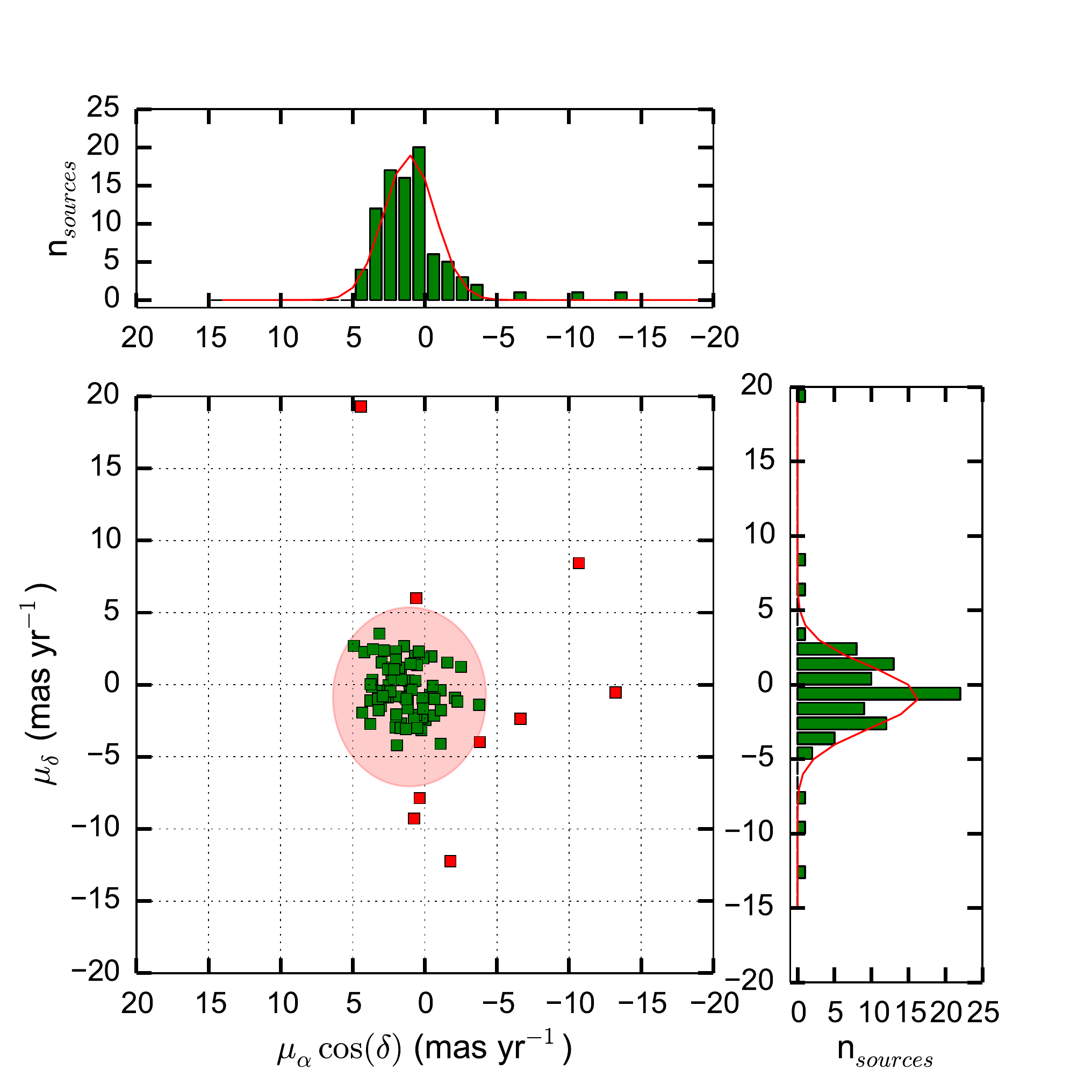}
\end{center}
\caption{Source distribution in the ($\mu_\alpha\cos(\delta),\mu_{\delta}$) plane.
Top and right panel shows the histogram distribution in right ascension and declination,
respectively, together with the Gaussian fits to the data. }
\label{fig:gau}
\end{figure*}

From the measured proper motion  we obtained the mean velocity
along both axes (right ascension and declination). This information is then used 
to identify stars with large peculiar motions.  As is shown in Figure~\ref{fig:gau},
stars with proper motions three or more times above the mean proper motion values,
red squares, are subsequently not used for our internal kinematic analysis
and are discussed separately. These stars are also shown as red arrows in
Figure~\ref{fig:pm}. 

For the remaining stars the mean proper motions are:

\begin{eqnarray*}
\overline{\mu_{\alpha}\cos{\delta}}=1.07\pm0.09\quad{\rm mas\,\,yr^{-1}},\\
\overline{\mu_{\delta}}=-0.84\pm0.16\quad{\rm mas\,\,yr^{-1}}.\\
\end{eqnarray*}

\noindent  Compared with those obtained by G\'omez et al. (2005), the mean proper 
motion in right ascension agrees within errors. On the other hand, the declination
value does not. As discussed in the previous section, the results of G\'omez et al. 
(2005) were affected by a systematic shift in this coordinate for the 1991.68
epoch (G\'omez et al. 2008). We attribute the differences to this shift. We argue 
that the values quoted above represent the bulk relative motion between the ONC and the Sun.

The transformation of the mean proper motions to the Galactic coordinate
system yields

\begin{eqnarray*}
\overline{\mu_{l}\cos{b}}=1.22\pm0.17\quad{\rm mas\,\,yr^{-1}},\\
\overline{\mu_{b}}=0.58\pm0.11\quad{\rm mas\,\,yr^{-1}}.\\
\end{eqnarray*}

\noindent The theoretical values of proper motions expected for a source in Orion,
can be derived from a Galactic rotation model and compared with our results. 
To estimate this expected motion, we assume, first that the stars move in circular 
orbits around the Galactic center
with an LSR speed of 254~km~s$^{-1}$ (Reid et al. 2009). Second, we adopt a Solar motion relative 
to the LSR of $(U,V,W)=(11.10, 12.24, 7.25)$~km~s$^{-1}$ (Sch\"onrich et al. 2010).
Finally, we assume that the Sun is at a distance of 8.4 kpc from the Galactic center (Reid et al. 2009).
From this Galactic rotation model the expected Galactic proper motions for stars 
at the position of Orion are:

\begin{eqnarray*}
\overline{\mu_{l}\cos{b}}\,(\rm exp)=1.39\pm0.05\quad{\rm mas\,\,yr^{-1}},\\
\overline{\mu_{b}}\,(\rm exp)=-0.02\pm0.02\quad{\rm mas\,\,yr^{-1}}.\\
\end{eqnarray*}

\noindent Thus, the average proper motion in Galactic longitude of  Orion is consistent with 
those expected from models of Galactic rotation. In Galactic latitude there is a peculiar
motion of 0.60$\pm0.12$~mas~yr$^{-1}\,\equiv\,1.18\pm0.24$~km~s$^{-1}$ toward the Galactic plane.

\begin{figure*}[ht!]
\begin{center}
\includegraphics[width=01.0\textwidth,trim= 50 0 40 30, clip]{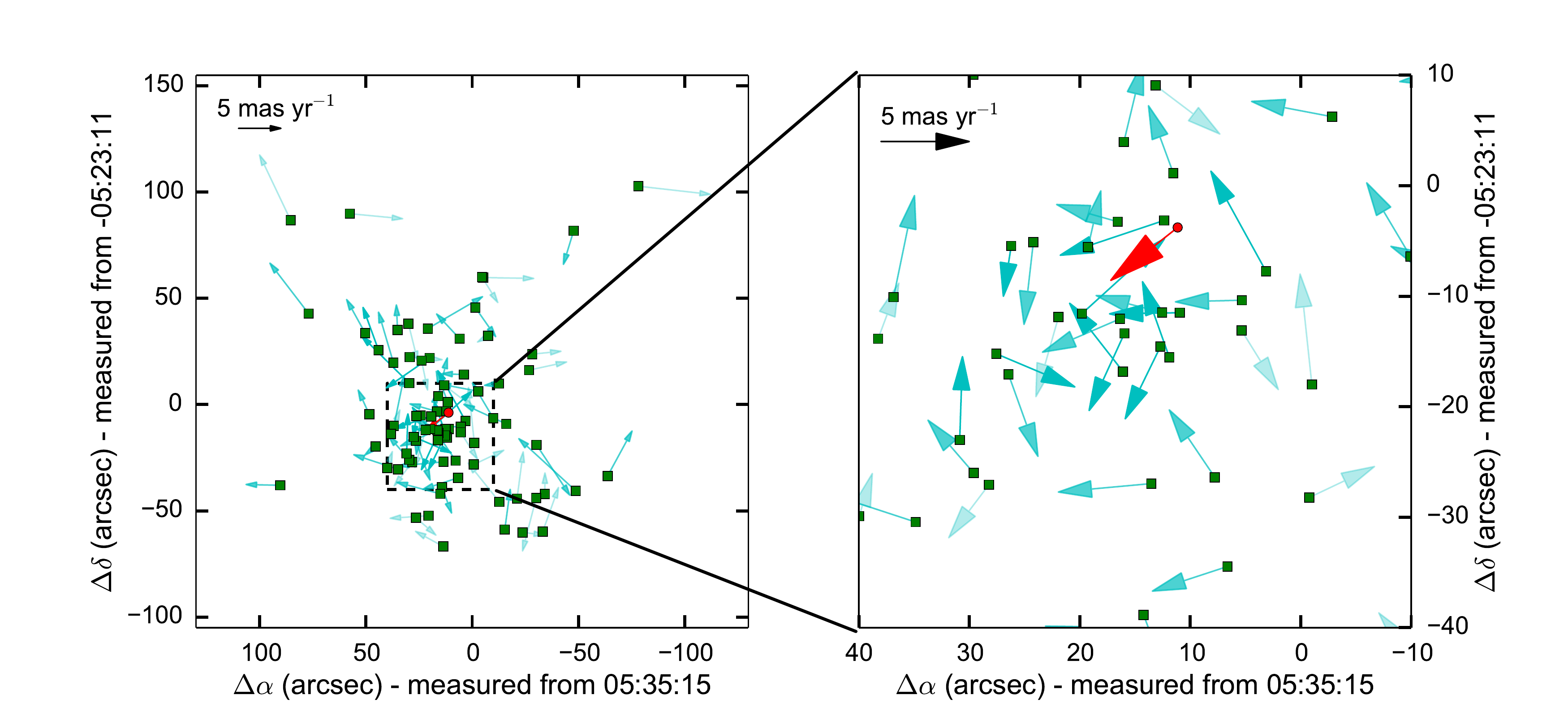}
\end{center}
\caption{Proper motions relative to the Orion's rest frame of radio stellar
sources with normal proper motions. The red arrow is the mean proper motion of the ONC.
{The level of transparency of the arrows indicates
the significance of the measured proper motions.
Left: the complete sample of relative proper motions.
Right: a zoom to the central region.}}
\label{fig:pmC}
\end{figure*}

Having measured the mean proper motion of the central ONC, we subtracted it from the individual
measured proper motions to calculate the proper motions of the sources in the rest frame of the 
ONC (Figure \ref{fig:pmC}).
The dispersion ($\sigma$) of these proper motions (after correction for the errors of measurement 
following Jones \& Walker 1988) are:

\begin{eqnarray*}
\sigma_{\alpha}=1.08\pm0.07\,{\rm mas\,\,yr^{-1}} \equiv2.12\pm0.13\,{\rm km\,\,s^{-1}},\\
\sigma_{\delta}=1.27\pm0.15\,{\rm mas\,\,yr^{-1}} \equiv2.49\pm0.29\,{\rm km\,\,s^{-1}}.\\
\end{eqnarray*}

The velocity dispersion values are lower than those found by G\'omez et al.~(2005)
for their radio sources analysis which are 
$\sigma_{\alpha,\,{\rm GRL2005}}=2.3\pm0.2\,{\rm mas\,yr^{-1}}$ and
$\sigma_{\delta,\,{\rm GRL2005}}=3.1\pm0.2\,{\rm mas\,yr^{-1}}$.
The differences in the results can be attributed to the different 
number of analyzed sources and to the fact that G\'omez et al. (2005) did not correct
for the errors of measurement. Our measured velocity dispersions
are in agreement with the velocity dispersions found by Jones \& Walker (1988) 
from optical observation of  
$\sigma_{\alpha,\,{\rm JW1998}}=0.91\pm0.06\,{\rm mas\,yr^{-1}}$ and
$\sigma_{\delta,\,{\rm JW1998}}=1.18\pm0.05\,{\rm mas\,yr^{-1}}$. 
Our computed velocity dispersions are also similar to the stellar radial 
velocity dispersion in the ONC of $\sigma_{V_{\rm rad}}\simeq2.5$~km~s$^{-1}$ 
(Kounkel et al.~2016b). Finally, they are also similar to the velocity 
dispersion on the plane of the sky of the ONC ionized gas of 
$\sigma_{pos}\simeq3\pm1$~km~s$^{-1}$, but significantly 
lower than the velocity dispersion along the line of sight of the ONC ionized 
gas $\sigma_{los}\simeq6\pm1$~km~s$^{-1}$ (Arthur et al. 2016). However, the 
latter could be affected by large velocity gradients and 
emissivity fluctuations along the line of sight (Arthur et al. 2016).

\subsection{The internal kinematics}

According to the review of Muench et al. (2008), there is some conflict
between results of studies of the kinematics of stars in Orion.
Some authors (e.g., Parenago 1954; Strand 1958; Fallon et al. 1977) have claimed
evidence of expansion or contraction, but it has also been
claimed that these results are due to observational errors 
(e.g., Vasilevskis 1962, 1971; Allen et al. 1974). Furthermore, given the
velocity dispersions, one would expect the ONC core to be virialized and not contracting
nor expanding motions are expected (e.g., Hillenbrand \& Hartmann 1998).

Using our measured proper motions, we searched for evidence of organized motions, 
specifically for expansion (or contraction) and rotation. 
First, we will use the proper motions of the radio sources relative to the Orion rest
frame that are listed in columns 5 and 6 of Table~\ref{tab:PM}. We follow
Rivera et al.~(2015) and define the vectors $r_*$, $\hat{r}_*$ and $\delta v_*$
for the position, the unit vector associated to the position, and the velocity 
for each star with respect to the center of the group. 
The cross products $\hat{r}_*\times\delta v_*$ and the dot products $\hat{r}_*\cdot \delta v_*$
are calculated individually for all the stars. The mean values of the cross and 
dot products can be used to search for organized movement.
In a purely radial movement the mean cross product is expected to be 
zero, while the mean dot product will be large (positive for expansion
and negative for contraction). On the other hand, for pure rotation 
the mean cross product will be large and the mean dot product is zero. 
For the YSOs in Orion we obtained:
\begin{eqnarray*}
\overline{\hat{r}_*\times\delta v_*}=0.7\pm0.3\quad{\rm km\,\,s^{-1}}{\rm ,\, and}\\
\overline{\hat{r}_*\cdot \delta v_*}=-0.1\pm0.3\quad{\rm km\,\,s^{-1}}.
\end{eqnarray*}
\noindent These numbers are small in comparison with the velocity dispersion, so our 
results do not point toward the existence of organized motions in the central ONC.
In particular there are no signs of expansion or contraction and this result agrees with 
previous discussions by Vasilevskis (1962, 1971) and Allen et al. (1974).

\subsection{Fast Moving Sources}

Stars with large peculiar velocities have been reported in the Orion 
Trapezium-BN/KL region in the past. The more interesting case is 
found in the BN/KL region where three stars (sources BN, I and the two
components related to source n) were suggested to 
participate in a dynamical decay event around 550 years ago (Rodr\'{\i}guez
et al. 2005, 2016; G\'omez et al. 2005, 2008), this event could also be related 
to the explosive phenomenon in the BN/KL region (Bally \& Zinnecker 2005). 
Two other stars were proposed to be ejected from the Orion Trapezium and 
one more was proposed to be ejected from the molecular core [TUK93]~28
a few thousand years ago (Poveda et al. 2005). The proposed mechanism that
produce these fast moving sources is $n$-body interactions (Poveda 
et al. 1967). We will use our results to identify new candidates 
to fast moving sources in this region.

From the 88 radio proper motion reported in this paper,
we found that, in addition to the four sources in the Orion BN/KL 
region, nine  present proper motion significantly
different from the rest of the radio sources in the core of the ONC (Figure \ref{fig:pm}). 
These sources are: V~1326~Ori, COUP~391, COUP~640, Parenago~1839,
MLLA~606, MLLA~630C, MLLA~712, Zapata~11 and IRc~23.  The measured
proper motions of COUP~640 and Parenago~1839 are below 1-$\sigma$, and thus
they  are not significant. We now discuss what could be 
the origin of the peculiar proper motions of the remaining fast moving sources.

First, we  check if these sources are foreground objects (e.g., Alves \& Bouy 2012).  
Sources at closer distances than Orion will have larger proper 
motions than sources that are actually part of this region.
From the Galactic rotation model described above we obtained the
expected proper motions of sources in the direction to Orion at 
different distances (Table \ref{tab:EqPMGR}). From this table 
we note that stars at closer distances will have larger and  more negative proper motion
in declination, while in right ascension the proper motion will be small and positive. 
Comparing these figures with the measured motions of our high peculiar proper motion 
sources, we conclude that  MLLA 606 and MLLA 630C 
could be sources at a closer distance to the Sun than the Orion system. 
Due to their large errors, however, it is not possible to favor any distance, and in fact
within errors these sources could also be part of the Orion system.

\begin{table}[!th]
\footnotesize
  \begin{center}
  \caption{Expected proper motions in the direction to Orion  from Galactic
  rotation model at different distances.}\label{tab:EqPMGR}
    \begin{tabular}{ccc}\hline\hline
Distance  &  $\mu_\alpha$  & $\mu_\delta$ \\
   (pc)       &    (mas)       & (mas)\\
   \noalign{\smallskip}
\hline\hline\noalign{\smallskip}
\noalign{\smallskip}
\hline\noalign{\smallskip}
100 &2.10$\pm$2.30    & -9.96$\pm$1.17  \\
200 &1.11$\pm$0.80    & \,-4.20$\pm$0.40  \\
300 &0.79$\pm$0.28    & \,-2.29$\pm$0.15    \\
400 &0.63$\pm$0.04    & \,-1.35$\pm$0.04 \\
 \hline\hline\noalign{\smallskip}
\end{tabular}
\end{center}
\end{table}

The proper motion vector of V 1326 Ori is one of the largest in our
sample. Interestingly, these motions suggest that in the past it
was closer to the densest region of the Trapezium group (see Figure \ref{fig:pm}). 
Thus, this source could have been ejected around 7,000 years 
ago from the Trapezium maybe also via $n$-body interactions. The proper motions
of this source in the rest frame of the ONC correspond to a linear velocity of
28.0$\pm$10.8 km s$^{-1}$.

\begin{figure}[ht!]
\begin{center}
\includegraphics[width=0.48\textwidth,trim= 20 50 0 0, clip]{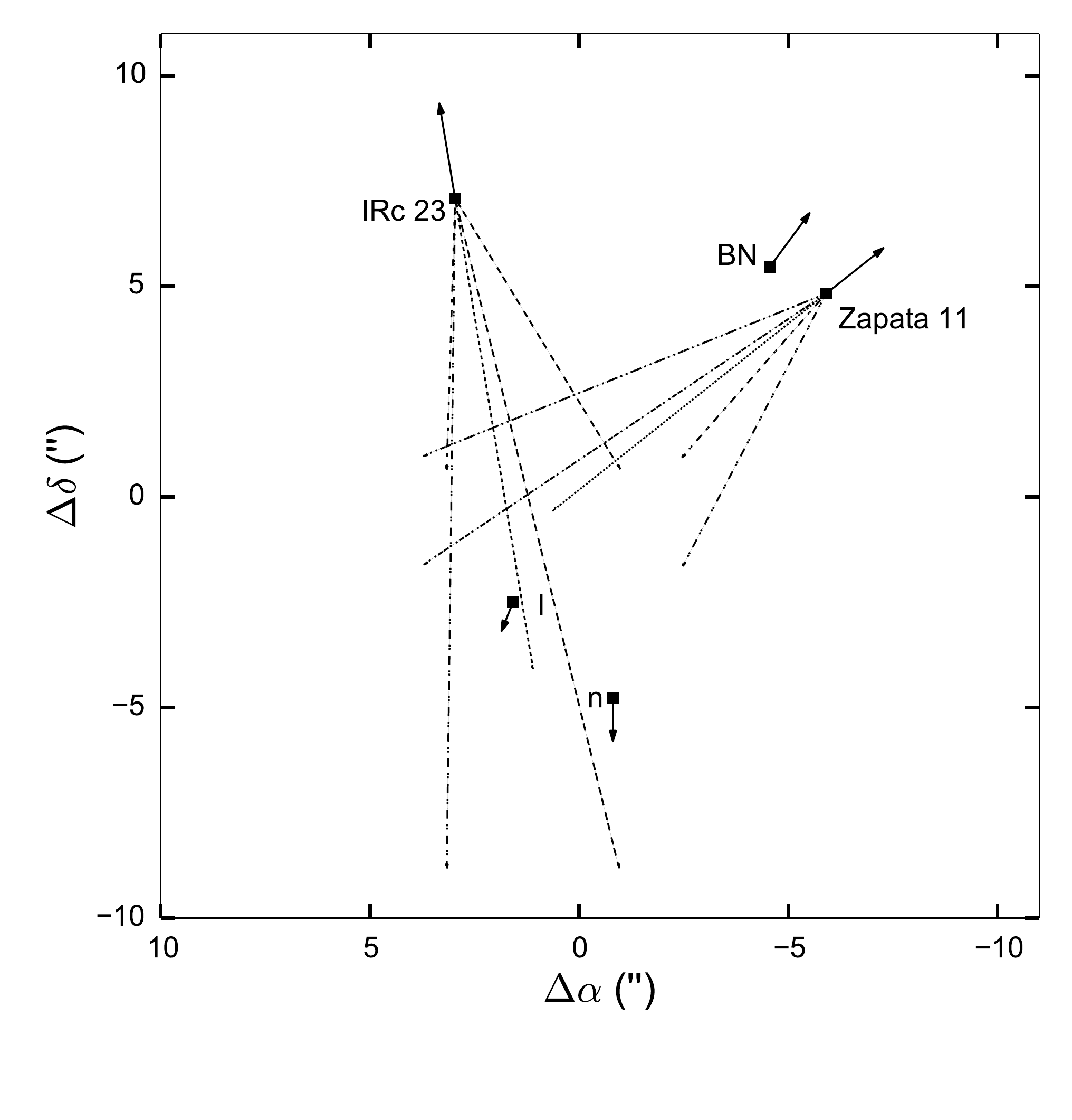}
\end{center}
\caption{Arrows indicate the direction and proper motion displacement for 100 years, 
in the rest frame of the Orion radio sources. The dotted  and dashed lines indicate the 
projected movement and errors, respectively, for the last 550 years of sources Zapata 11 
and IRc 23. 
Proper
motions of sources BN, I and n were taken from Rodr\'{\i}guez et al. (2016). The origin 
coordinates are RA=\rahms{05}{35}{14}{41}; dec.=\decdms{-05}{22}{28}{1};  { the mean position
of sources BN, I and n on their closest approach around 550 years ago} 
(Rodr\'{\i}guez et al. 2016).}
\label{fig:bnkl}
\end{figure}

The source Zapata 11 is $1\rlap{.}''5$ southwest from the famous BN source
and it was first noticed by Menten \& Reid (1995), who suggested that it is a jetlike
extension of this source. The measured proper motion of Zapata 11 are comparable 
with those of BN (G\'omez et al. 2008; Rodr\'{\i}guez et al. 2016; see also 
Figure~\ref{fig:bnkl}) and there is no significant evidence that it is moving away 
rapidly from BN. Thus Zapata~11 could be, instead, a companion of BN, although 
the separation between the two ($\sim$621 AU) seems somewhat large for sources in the 
region (for example see the discussion by Petr et al.~1998). 
Additional observations could help to distinguish between these
two hypothesis. On the other hand, source IRc~23 is located  $\sim7\rlap{.}''5$ northeast of
BN. It was first reported at radio frequencies by Forbrich et al. (2016) 
\footnote{ It was, however, clearly detected before by Menten \& Reid (1995) and by G\'omez et al. (2008).
These last authors mislabeled it as source D in their Figure 2.}
with a
spectral index that suggests non-thermal emission. Extrapolating the proper motions
of IRc 23 to 550 years in the past it appears to agree, within the errors, with the position where 
sources BN, I and n were then (Figure \ref{fig:bnkl}). However, due to the large uncertainties, the relation of IRc 23 with
the other sources in BN/KL needs to be investigated further.

Finally, the sources COUP 391 and MLLA 712 have proper motions with
large uncertainties, and within errors they are also 
consistent within three times the mean proper motion, our criteria to select
fast moving sources. Future observations
could help to decide if these objects have large peculiar motions or not. 

\section{CONCLUSIONS}

Using observations with the VLA radio interferometer that span 29.1 years, 
we measured the absolute proper motions of 88 YSOs with radio emission in the Orion 
Trapezium-BN/KL region. The analysis of these proper motions let us reach the 
following conclusions:

1. The mean proper motions of the Orion Trapezium-BN/KL regions are 
$\overline{\mu_{\alpha}\cos{\delta}}=1.03\pm0.10\quad{\rm mas\,yr^{-1}}$ and
$\overline{\mu_{\delta}}=-0.82\pm0.16\quad{\rm mas\,yr^{-1}}$. These proper motions
agree within errors with those expected from Galactic rotation curves, with only a small
peculiar motion of 1.14$\pm0.24$~km~s$^{-1}$ toward the
Galactic plane.

2. The calculated velocity dispersions are $\sigma_{\alpha}=2.12\pm0.13\quad{\rm km\,\,s^{-1}},$
$\sigma_{\delta}=2.49\pm0.29\quad{\rm km\,\,s^{-1}}$. These values are in agreement with those obtained
at optical wavelengths from proper motions by Jones \& Walker (1988), stellar radial velocities 
by Kounkel et al. (2016b), { and from gas radial velocities by Arthur et al. (2016)}.

3. The kinematics of the sources do not show evidence for expansion, contraction
or rotation of the ONC.

4. The large proper motions measured for V 1326 Ori indicate that it could be
a runaway star, and the proper motion vector suggests that it might have been ejected
from the Orion Trapezium around 7,000 years ago.

5. Extrapolating the proper motions of radio sources Zapata 11 and IRc 23 to 550 years in the past,
we find that their positions  are close to the intersecting position of sources BN, I and n, the massive
stars that participated in a dynamical disintegration. Source Zapata 11 could be either a
companion or a slow ejecta of BN. 

6. The proper motions of six other radio sources also indicate large peculiar motions.
However, within the errors they could also follow the bulk kinematics of the other sources
in the region. For two of them, however, the proper motions could be more naturally explained if
they were objects along the line of sight to, but at a smaller distance than, the ONC.

The astrometric radio studies of the Orion Trapezium-BN/KL region appears promising 
for the future, since Forbrich et al. (2016) detected a total of 556 compact radio
sources in this region. With additional new high resolution and high sensitivity 
radio observations distributed in the next decade, it will be possible to
measure the proper motion of these hundreds of sources and perform a comprehensive
internal kinematic study of YSOs in the densest region of Orion.

\acknowledgments
L.L., L.F.R., and J.L.R. acknowledge the financial 
support of DGAPA, UNAM, and CONACyT, M\'exico. 
The National Radio Astronomy Observatory 
is operated by Associated Universities Inc. under
cooperative agreement with the National 
Science Foundation. This research has made use of 
the SIMBAD database, operated at CDS, Strasbourg, France

\appendix

\section{Improvement of the errors in proper motion determinations as a function of time}

The least-squares fitting to the positions of a source as a function of time is a linear regression problem.
Following Press et al. (1992), the position $x$ in either right ascension or declination is described as a function
of time $t$ as:

\begin{equation}
  x = a + b t\,,
  \label{xy}
\end{equation}
where $a$ is the intercept of the line with the x-axis and $b$ is the slope. If we have a set of $N$ data points ($t_i$, $x_i$)
we can define the following sums:

\begin{equation}
S = \sum_{i=0}^{N-1}{{1}\over{\sigma_i^2}} ~~~~~~~~~~~~~~~  
S_t = \sum_{i=0}^{N-1}{{t_i}\over{\sigma_i^2}}~~~~~~~~~~~~~~~
S_x = \sum_{i=0}^{N-1}{{x_i}\over{\sigma_i^2}}~~~~~~~~~~~~~~~
\end{equation}
\begin{equation}
S_{tt} = \sum_{i=0}^{N-1}{{t_i^2}\over{\sigma_i^2}}~~~~~~~~~~~~~~~
S_{tx} = \sum_{i=0}^{N-1}{{t_i x_i}\over{\sigma_i^2}}\,. 
\label{sum}
\end{equation}

In these sums $\sigma_i$ is the error associated to each $x_i$ value. Additionally defining

\begin{equation}
\Delta = S S_{tt} - (S_t)^2\,,
\label{de}
\end{equation}
the best fit model parameters are given by:

\begin{equation}
a = {{S_{tt} S_x - S_t S_{tx}}\over{\Delta}}\,,
\label{aa}
\end{equation}
\begin{equation}
b = {{S S_{tx} - S_t S_x}\over{\Delta}}\,.
\label{bb}
\end{equation}

Finally, the errors of the parameters are given by:

\begin{equation}
\sigma_a = \sqrt{{{S_{tt}}\over{\Delta}}}\,,
\label{asigma}
\end{equation}
\begin{equation}
\sigma_b = \sqrt{{{S}\over{\Delta}}}\,.
\label{bsigma}
\end{equation}

To simplify the calculations we will assume that the measurements are made evenly in time:

\begin{equation}
t = 0, \Delta t, 2 \Delta t, 3 \Delta t,...,(N-1)\Delta t\,,
\end{equation}
where $\Delta t$ is the time interval between measurements.
We also assume that the error of each measurement is the same:

\begin{equation}
\sigma_i = \sigma\,.
\end{equation}

With these simplifications, three of the sums listed above are given by:

\begin{equation}
S = \sum_{i=0}^{N-1}{{1}\over{\sigma_i^2}} = {{N}\over{\sigma^2}}
\end{equation}
\begin{equation}
S_t = \sum_{i=0}^{N-1}{{t_i}\over{\sigma_i^2}} = {{N (N-1)}\over{2 \sigma^2}} \Delta t 
\end{equation}
\begin{equation}
S_{tt} = \sum_{i=0}^{N-1}{{t_i^2}\over{\sigma_i^2}} = {{(2N-1)(N-1)(N)}\over{6 \sigma^2}} (\Delta t)^2 
\end{equation}
  
In equations (A12) and (A13) we have used the formulae for the addition of the natural 
numbers and the addition of the squares of
the natural numbers, respectively. Equation (A4) can be approximated by:

\begin{equation}
\Delta = S S_{tt} - (S_t)^2 = {{N^2(N^2 - 1)}\over{12 \sigma^4}}(\Delta t)^2  \simeq {{N^4}\over{12 \sigma^4}} (\Delta t)^2\,,
\end{equation}
where we have assumed that $N^2 \gg 1$. Finally, substituting in equations (A7) and (A8) we obtain:

\begin{equation}
\sigma_a = \sqrt{{{S_{tt}}\over{\Delta}}} = \sigma \sqrt{{{4}\over{N}}}\,,
\end{equation}
\begin{equation}
\sigma_b = \sqrt{{{S}\over{\Delta}}} = {{\sigma}\over{\Delta t}} \sqrt{{{12}\over{N^3}}}\,.
\end{equation}

In summary, the error in the intercept will decrease as $time^{-1/2}$, while the error in the slope
will decrease as $time^{-3/2}$. Thus, duplicating the time coverage will improve the determination of 
proper motions by an important factor of 2.8.


\begin{thebibliography}{}

\bibitem[Allen et al.(1974)]{1974RMxAA...1..101A} Allen, C., Poveda, A., \& Worley, C.~E.\ 1974, \rmxaa, 1, 101

\bibitem[Alves \& Bouy(2012)]{2012A&A...547A..97A} Alves, J., \& Bouy, H.\ 2012, \aap, 547, A97 

\bibitem[Arthur et al.(2016)]{2016MNRAS.463.2864A} Arthur, S.~J., Medina, S.-N.~X., \& Henney, W.~J.\ 2016, \mnras, 463, 2864

\bibitem[Bally et al.(2000)]{2000AJ....119.2919B} Bally, J., O'Dell, C.~R., \& McCaughrean, M.~J.\ 2000, \aj, 119, 2919 

\bibitem[Bally \& Zinnecker(2005)]{2005AJ....129.2281B} Bally, J., \& Zinnecker, H.\ 2005, \aj, 129, 2281 

\bibitem[Churchwell et al.(1987)]{1987ApJ...321..516C} Churchwell, E., Felli, M., Wood, D.~O.~S., \& Massi, M.\ 1987, \apj, 321, 516 

\bibitem[Dzib et al.(2014)]{2014ApJ...788..162D} Dzib, S.~A., Loinard, L., 
Rodr{\'{\i}}guez, L.~F., \& Galli, P.\ 2014, \apj, 788, 162 

\bibitem[Fallon et al.(1977)]{1977ApJ...217..719F} Fallon, F.~W., Gerola, H., \& Sofia, S.\ 1977, \apj, 217, 719 

\bibitem[Forbrich et al.(2016)]{2016ApJ...822...93F} Forbrich, J., Rivilla, V.~M., Menten, K.~M., et al.\ 2016, \apj, 822, 93 

\bibitem[Garay et al.(1987)]{1987ApJ...314..535G} Garay, G., Moran, J.~M., \& Reid, M.~J.\ 1987, \apj, 314, 535

\bibitem[Getman et al.(2005)]{2005ApJS..160..319G} Getman, K.~V., Flaccomio, E., Broos, P.~S., et al.\ 2005, \apjs, 160, 319 

\bibitem[G{\'o}mez et al.(2005)]{2005ApJ...635.1166G} G{\'o}mez, L., 
Rodr{\'{\i}}guez, L.~F., Loinard, L., et al.\ 2005, \apj, 635, 1166 

\bibitem[G{\'o}mez et al.(2008)]{2008ApJ...685..333G} G{\'o}mez, L., 
Rodr{\'{\i}}guez, L.~F., Loinard, L., et al.\ 2008, \apj, 685, 333 

\bibitem[Hillenbrand(1997)]{1997AJ....113.1733H} Hillenbrand, L.~A.\ 1997, \aj, 113, 1733 


\bibitem[Hillenbrand \& Hartmann(1998)]{1998ApJ...492..540H} Hillenbrand, L.~A., \& Hartmann, L.~W.\ 1998, \apj, 492, 540 

\bibitem[Jones \& Walker(1988)]{1988AJ.....95.1755J} Jones, B.~F., \& Walker, M.~F.\ 1988, \aj, 95, 1755

\bibitem[Kounkel et al.(2014)]{2014ApJ...790...49K} Kounkel, M., Hartmann, 
L., Loinard, L., et al.\ 2014, \apj, 790, 49 

\bibitem[Kounkel et al.(2016a)]{2016Kounkel} Kounkel, M., Hartmann, 
L., Loinard, L., et al.\ 2016, submitted to \apj

\bibitem[Kounkel et al.(2016b)]{2016ApJ...821....8K} Kounkel, M., Hartmann, L., Tobin, J.~J., et al.\ 2016, \apj, 821, 8 

\bibitem[Kukarkin et al.(1971)]{1971GCVS3.C......0K} Kukarkin, B.~V., Kholopov, P.~N., Pskovsky, Y.~P., et al.\ 1971, General Catalogue of Variable Stars, 3rd ed.~(1971),  

\bibitem[Loinard(2002)]{2002RMxAA..38...61L} Loinard, L.\ 2002, \rmxaa, 38, 
61 

\bibitem[Loinard et al.(2007)]{2007ApJ...671..546L} Loinard, L., Torres, R.~M., Mioduszewski, A.~J., et al.\ 2007, \apj, 671, 546 


\bibitem[McMullin et al.(2007)]{2007ASPC..376..127M} McMullin, J.~P., Waters, B., Schiebel, D., Young, W., \& Golap, K.\ 2007, Astronomical Data Analysis Software and Systems XVI, 376, 127 

\bibitem[Menten \& Reid(1995)]{1995ApJ...445L.157M} Menten, K.~M., \& Reid, M.~J.\ 1995, \apjl, 445, L157 

\bibitem[Menten et 
al.(2007)]{2007A&A...474..515M} Menten, K.~M., Reid, M.~J., Forbrich, J., \& Brunthaler, A.\ 2007, \aap, 474, 515 

\bibitem[Muench et al.(2008)]{2008hsf1.book..483M} Muench, A., Getman, K., Hillenbrand, L., \& Preibisch, T.\ 2008, Handbook of Star Forming Regions, Volume I, 4, 483 

\bibitem[Muench et al.(2002)]{2002ApJ...573..366M} Muench, A.~A., Lada, E.~A., Lada, C.~J., \& Alves, J.\ 2002, \apj, 573, 366

\bibitem[O'Dell et al.(2008)]{2008hsf1.book..544O} O'Dell, C.~R., Muench, A., Smith, N., \& Zapata, L.\ 2008, Handbook of Star Forming Regions, Volume I, 4, 544 

\bibitem[Ortiz-Le\'on et al.(2016)]{2016Ortiz} Ortiz-Le\'on, G. N., Loinard, L., Dzib, S.~A., et al.\ 2016, submitted to \apj

\bibitem[Parenago(1954)]{1954TrSht..25....1P} Parenago, P.~P.\ 1954, Trudy Gosudarstvennogo Astronomicheskogo Instituta, 25, 393 

\bibitem[Petr et al.(1998)]{1998ApJ...500..825P} Petr, M.~G., Coud{\'e} du Foresto, V., Beckwith, S.~V.~W., Richichi, A., \& McCaughrean, M.~J.\ 1998, \apj, 500, 825 

\bibitem[Poveda et al.(1967)]{1967BOTT....4...86P} Poveda, A., Ruiz, J., \& Allen, C.\ 1967, Boletin de los Observatorios Tonantzintla y Tacubaya, 4, 86 

\bibitem[Poveda et al.(2005)]{2005ApJ...627L..61P} Poveda, A., Allen, C., \& Hern{\'a}ndez-Alc{\'a}ntara, A.\ 2005, \apjl, 627, L61

\bibitem[Pradel et al.(2006)]{2006A&A...452.1099P} Pradel, N., Charlot, P., \& Lestrade, J.-F.\ 2006, \aap, 452, 1099

\bibitem[Press(1992)]{Press1992} Press, W.~H., Teukolsky, S.~A., Vetterling, W.~T., \& Flannery, B.~P.\ 1992, 
Numerical recipes in FORTRAN. The art of scientific computing, Cambridge: University Press, 2nd ed.  

\bibitem[Reid et al.(2009)]{2009ApJ...700..137R} Reid, M.~J., Menten, K.~M., Zheng, X.~W., et al.\ 2009, \apj, 700, 137-148 

\bibitem[Rieke et al.(1973)]{1973ApJ...186L...7R} Rieke, G.~H., Low, F.~J., \& Kleinmann, D.~E.\ 1973, \apjl, 186, L7 

\bibitem[Rivera et al.(2015)]{2015ApJ...807..119R} Rivera, J.~L., Loinard, 
L., Dzib, S.~A., et al.\ 2015, \apj, 807, 119 

\bibitem[Rodr{\'{\i}}guez et al.(2003)]{2003ApJ...583..330R} 
Rodr{\'{\i}}guez, L.~F., Curiel, S., Cant{\'o}, J., et al.\ 2003, \apj, 
583, 330 

\bibitem[Rodr{\'{\i}}guez et al.(2005)]{2005ApJ...627L..65R} Rodr{\'{\i}}guez, L.~F., Poveda, A., Lizano, S., \& Allen, C.\ 2005, \apjl, 627, L65 

\bibitem[Rodr{\'{\i}}guez et al.(2016)]{2016submittedApJ} Rodr{\'{\i}}guez, L.~F., Dzib, S.~A., Loinard, L., et al. 2016, submitted to ApJ. 

\bibitem[Sch{\"o}nrich et al.(2010)]{2010MNRAS.403.1829S} Sch{\"o}nrich, R., Binney, J., \& Dehnen, W.\ 2010, \mnras, 403, 1829 

\bibitem[Strand(1958)]{1958ApJ...128...14S} Strand, K.~A.\ 1958, \apj, 128, 14 



\bibitem[Vasilevskis(1962)]{1962AJ.....67..699V} Vasilevskis, S.\ 1962, \aj, 67, 699 

\bibitem[Vasilevskis(1971)]{1971ApJ...167..537V} Vasilevskis, S.\ 1971, \apj, 167, 537 


\bibitem[Zapata et al.(2004)]{2004AJ....127.2252Z} Zapata, L.~A., 
Rodr{\'{\i}}guez, L.~F., Kurtz, S.~E., 
\& O'Dell, C.~R.\ 2004, \aj, 127, 2252 

\end{thebibliography}
\end{document}